\documentclass[pdflatex,sn-mathphys-num]{sn-jnl}

\usepackage{graphicx}%
\usepackage{multirow}%
\usepackage{amsmath,amssymb,amsfonts}%
\usepackage{amsthm}%
\usepackage{mathrsfs}%
\usepackage[title]{appendix}%
\usepackage{xcolor}%
\usepackage{textcomp}%
\usepackage{manyfoot}%
\usepackage{booktabs}%
\usepackage{algorithm}%
\usepackage{algorithmicx}%
\usepackage{algpseudocode}%
\usepackage{listings}%
\usepackage{siunitx}

\theoremstyle{thmstyleone}%
\theoremstyle{thmstyletwo}%

\theoremstyle{thmstylethree}%

\begin{document}

\title[Article Title]{Nonlinear Dynamics and Fermi–Pasta–Ulam–Tsingou Recurrences in Macroscopic Ultra-low Loss Levitation}

\author[1]{\fnm{Mehrdad M.} \sur{Sourki}}

\author[1]{\fnm{Wisdom} \sur{Boinde}}

\author[1]{\fnm{Ali N.} \sur{Amiri}}

\author*[1,2]{\fnm{Mahdi} \sur{Hosseini}}\email{mh@northwestern.edu}

\affil[1]{\orgdiv{Department of Electrical and Computer Engineering}, \orgname{Northwestern University}, \orgaddress{\street{2145 Sheridan Rd}, \city{Evanston}, \postcode{60028}, \state{IL}, \country{U.S.A}}}
\affil[2]{\orgdiv{Applied Physics Program}, \orgname{Northwestern University}, \orgaddress{\street{2145 Sheridan Rd}, \city{Evanston}, \postcode{60028}, \state{IL}, \country{U.S.A}}}


\abstract{Macroscopic systems, when governed by nonlinear interactions, can display rich behavior from persistent oscillations to signatures of ergodicity breaking. Nonlinearity, long regarded as a nuisance in precision systems, is increasingly recognized as a gateway to new physical regimes. While such dynamics have been extensively studied in optics and atomic physics, macroscopic systems are rarely associated with long-lived coherence and nonlinear control  and remain an untapped platform for probing the fundamental nonlinear processes. Here, we report the observation of long-lived oscillatory dynamics in millimeter-scale levitated dielectric quartz particles exhibiting clear signatures of nonlinear mode coupling, a positive largest Lyapunov exponent of $0.0095$~s$^{-1}$, and partial energy recurrences—phenomena strongly reminiscent of the Fermi–Pasta–Ulam–Tsingou physics.  We observe dissipation rates below $4\times10^{-6}~ \text{Hz}$, limited by our ability to measure dissipation in presence of nonlinear dynamics.  We estimate an intrinsic acceleration sensitivity of $62\times10^{-12}~ g/\sqrt{\text{Hz}}$, at room temperature. The magnetic trap is constructed from a static arrangement of permanent magnets, requiring no external power or active feedback. Our findings open a path toward leveraging nonlinear dynamics for novel applications in sensing, signal processing, and statistical mechanics.}


\maketitle

\maketitle

\section*{Main}\label{sec1}
The study of nonlinear dynamics across various physical systems has profoundly transformed our understanding of fundamental processes and enabled important applications.  The boundary between stability and chaos in nonlinear dynamical systems is far more intricate than initially envisioned. A paradigmatic example is the Fermi–Pasta–Ulam–Tsingou (FPUT) problem \cite{osti_4376203, midtvedt_fermi-pasta-ulam_2014}, where instead of rapid thermalization, the system exhibits long-lived recurrences and quasi-integrable dynamics, even at energies where chaos might have been expected. Despite decades of investigation, the precise mechanisms governing the onset of chaos, the conditions under which recurrences persist, and the generality of these features across different physical platforms remain unresolved. These open questions have far-reaching implications: they challenge our understanding of thermalization in isolated systems, test the limits of predictability, and impact fields ranging from condensed matter physics to quantum information science. 
In optical systems, investigations into nonlinear wave propagation and soliton formation have directly led to the development of optical frequency combs \cite{fortier201920, ctx44402003590002441}, which are now essential tools in precision metrology \cite{RevModPhys.78.1297}, spectroscopy \cite{Picqu__2019}, and timekeeping \cite{giorgetta2013optical,deschenes2016synchronization,caldwell2022time}. In quantum systems, nonlinear interactions underpin phenomena such as photon blockade \cite{PhysRevLett.79.1467}, quantum squeezing \cite{
PhysRevLett.55.2409}, and many-body effects \cite{Pizzi_2025} with no classical analogs.

Mechanical oscillators can also exhibit strong nonlinearity, leading to rich dynamical behavior and routes to thermalization that differ significantly from their linear counterparts. However, nonlinear mechanical systems have not been as extensively explored, in part due to the technical challenges associated with precisely engineering and measuring nonlinear potentials, especially in macroscopic or low-dissipation regimes. Recent advances in levitation-based platforms \cite{doi:10.1126/science.aba3993, PhysRevLett.105.101101,leng_mechanical_2021}  and micro- and nanoscale resonators \cite{PhysRevLett.116.147202, Verhagen_2012} offer new opportunities to probe these systems, enabling controlled studies of nonlinear physics in both classical and quantum settings.

Diamagnetic levitation of macroscopic objects has recently emerged as a promising platform for precision sensing \cite{xiong_achievement_2025,chen2025levitatedmacroscopicrotors10}, quantum-limited measurements \cite{PhysRevLett.109.147206}, and experimental tests of fundamental theories \cite{PhysRevResearch.2.013057, PhysRevResearch.2.043229}. Diamagnetic potentials offer stable three-dimensional trapping of macroscopic objects with no external drive, no feedback, and no contact loss channels, even at room temperature.  Composite graphite materials \cite{chen_diamagnetic_2022,tian_feedback_2024, chen2025levitatedmacroscopicrotors10} and sub-nanogram dielectric materials \cite{leng_mechanical_2021} have been levitated exhibiting very low dissipation rates. While previous work has focused on nanogram-scale dielectric particles or large non-dielectric materials such as composite graphene, the significantly larger mass and size of levitated object can benefit from inertial stability and reduced noise and dissipation.  Furthermore, diamagnetically levitated particles naturally exhibit nonlinearities \cite{ chen2024nonlinear,PhysRevResearch.4.013132} due to the anharmonic trapping potential and geometric asymmetries. Thus levitating macroscopic dielectric particles at the millimeter scale opens a new regime in the developing novel sensors as well as the study of nonlinear and nonequilibrium dynamics.

Despite their conceptual simplicity, such levitated macroscopic systems have remained largely unexplored as platforms for studying thermalization pathways. Our system comprises a millimeter-scale dielectric particle of asymmetric shapes levitated above a permanent magnet array. We observe and characterize multiple underdamped vibrational modes, with lifetimes $>10^4$ seconds, and track their long-time evolution under small perturbations. The dynamics exhibit hallmark features of FPUT problem , including persistent nonlinear intermodal coupling, long-lived energy recurrences, incomplete thermalization over experimentally accessible timescales, and coherent higher harmonic generation.
 

Diamagnetic levitation is the stable suspension of an object in a magnetic field due to its diamagnetic response. Diamagnetism arises from the induced magnetic moments in materials that oppose the applied magnetic field, resulting in a repulsive force.

For a diamagnetic material with volume \( V \), magnetic susceptibility \( \chi < 0 \), and permeability of free space \( \mu_0 \), the magnetic potential energy is: $U = -\frac{\chi V}{2\mu_0} |\vec{B}|^2$. Stable levitation is achieved when the corresponding magnetic force balances gravity and provides a restoring force in all directions. For a particle of mass \( m \) under gravity \( g \), the vertical equilibrium condition becomes: $\frac{\chi V}{\mu_0} \left( \frac{dB^2}{dz} \right) = mg$
The trap frequency \( \omega_0 \) is determined by the curvature of the magnetic field at the equilibrium position. Given that the density and magnetic susceptibility are different from one material to another, the trap can levitate different materials with different levels of ease and at certain equilibrium heights. Table \ref{table1} shows the density and magnetic susceptibility of several materials and the levitation parameter ($B\frac{dB}{dz}$), which corresponds to the field strength and its gradient needed to trap various materials. A larger levitation parameter corresponds to increased difficulty in achieving stable trapping. Another factor in choosing the levitation material is low conductivity to reduce eddy-current damping. As it can be seen, quartz and N-BK7 have negligible conductivity but are among the most challenging materials to stably levitate diamagnetically. Using a tailored configuration of permanent magnets, we successfully levitated macroscopic (mm-scale) quartz and N-BK7 objects, attaining ultra-low dissipation rates that reveal the intrinsic nonlinear dynamics of the trapping potential.

\begin{table}
    \centering
    \begin{tabular}{c c c c c}
        Material & Density  & Magnetic  &  $B \frac{\partial B }{ \partial z} $ & Electrical \\
         &  $\rho[gr/cm^3]$ &Susceptibility&$[T/m^2]$ & Conductivity\\
          &   &$\chi$ $(-10^{-5})$ & & $\sigma (S/m)$\\\hline\hline
        Pyrolytic graphite  &  2.3 & 45 \cite{simon2000diamagnetic} & 61 & $2.5 \times 10^{6}$ (in Plane) \\\hline
      Bismuth  &  9.8 & 16 \cite{haynes2016crc} & 730 & $9.3 \times 10^{5}$ \\\hline
      DI Water   &  1.0 & 0.9 \cite{simon2000diamagnetic} & 1400 & $ \mathcal{O}(10^{-6})$ \\\hline
      Quartz   &  2.6 & 1.4 \cite{hrouda1986effect} & 2000 & $ \mathcal{O}(10^{-18})$\\\hline
Diamond  &  3.5 & 2.2 \cite{haynes2016crc}  & 2000& $\mathcal{O}(10^{-14})$\\\hline
      N-BK7  &  2.5 & 1.2 \cite{wapler2014magnetic} & 2700 &$\mathcal{O} ( 10^{-8})$\\
    \end{tabular}
    \caption{Comparison of  relevant diamagnetic materials}
    \label{table1}
\end{table}

\begin{figure}
    \centering
    \includegraphics[width=1\linewidth]{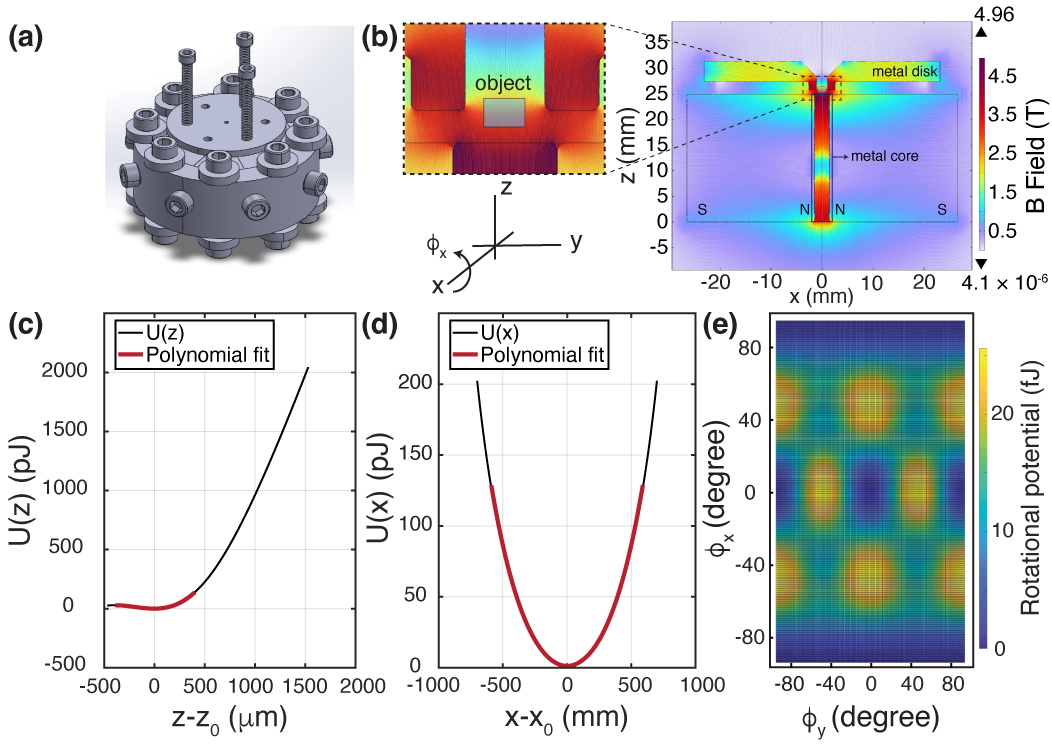}
    \caption{(a) 3D illustration of magnetic assembly consisting of eight triangle-shaped magnets held together to create a radial magnet. A metallic core and a metallic disk are used to focus the field at the center above the magnet. (b) COMSOL field simulation shown in 2D for the whole magnet assembly and a closed up view near the levitation point. A cylindrical opening (about 2~mm in diameter) at the center of the top metallic disk guides the magnetic flux from the metal core and provides space for levitation. (c)-(e) Sum of magnetic and gravitational potentials integrated over a cubic volume made of quartz  (side=0.5~mm, mass$\simeq$0.3~mg) plotted for coordinates $x$, $z$ and angular displacement around $x$ and $y$ axes ($\phi_x~\&~\phi_y$), respectively.}
    \label{fig1}
\end{figure}

\section*{Experimental Results}
A set of N42 neodymium magnets are assembled as shown in Fig.\ref{fig1} (a)-(b) to focus the magnetic field in a mm-region above the magnet. A metallic rod and disk enable focusing the field in the gap above the center of the magnets where a strong field gradient enables 3D diamagnetic trapping of mm-scale dielectrics. Using this trap, we have achieved stable levitation of different weakly diamagnetic dielectrics (silica, NB-K7 glass, and crystalline quartz) of symmetric and asymmetric shapes (sphere, hemisphere, cylinder, and cube).

Fig.~\ref{fig1} (c)-(e) show the total magnetic and gravitational potential calculated for a cubic volume of quartz (side=0.5mm, mass$\simeq$0.3~mg). In Fig.~\ref{fig1} (c) the potential is calculated as the position of the center of mass (CoM) of the cube has been moved along the $z$ axis starting from $\SI{250}{\micro\meter}$ above the magnets. As this simulation shows, this potential has a minima at $\sim \SI{850}{\micro\meter} $ above the magnets. This agrees with our experimental observation. The simulated trap along this axis shows a stiffness of about $1.2\times 10^{-3}$ N/m and an associated frequency of about $10.3$~Hz. Fig.~\ref{fig1} (d) shows the trap potential in lateral directions ($x$ or $y$). The trap frequency and stiffness along the horizontal plane are $7.1$~Hz and $5.5\times 10^{-4}$~N/m respectively. Fig.~\ref{fig1} (e) shows the effect of the rotation of the cube around $x$ and $y$ axis and its ensuing change in potential. As this figure shows, the stable position is when the cube is sitting with its face parallel to the horizontal plane. Depending on the rotation around $x$ and $y$, this two-dimensional trap can have different depths along the $y$ and $x$ axes, leading its rotational motion to assume frequencies ranging from $0$~Hz up to $\sim0.6$~Hz near the stable arrangement. Also, it is worth noting that the rotation around the $z$ axis is only constrained by the trap imperfections leading to field asymmetries around the vertical axis. As can be seen below, these predictions are in good agreement with the experimental observations.

\begin{figure}
    \centering
    \includegraphics[width=0.8\linewidth]{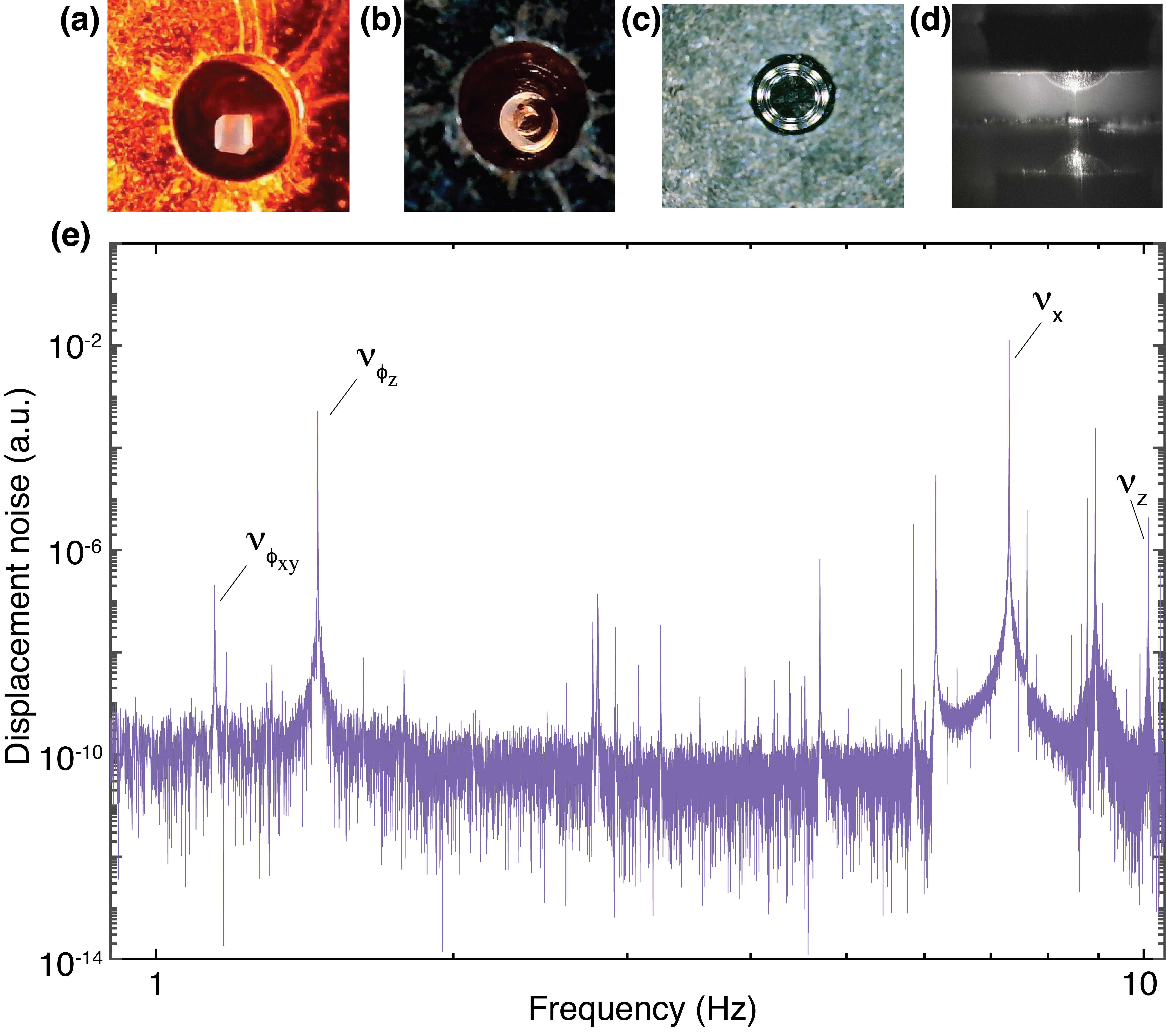}
    \caption{(a)-(c) Top view images of levitated quartz cube (side=0.5~mm, mass$\simeq0.3$~mg), hollow quartz cylinder (diameter= 0.5~mm, mass$\simeq2$~mg), and N-BK7 hemisphere (diameter=1.5~mm, mass$\simeq$2.5~mg), respectively. (d) Side view image of levitated hemisphere. (e) A typical vibrational spectrum for levitated cube measured using non-interferometric single-pixel detection, with main vibrational modes indicated.}
    \label{fig2}
\end{figure}

The levitation setup is placed inside a vacuum chamber reaching pressures as low as $10^{-8}$ Torr. Images of levitated objects are shown in Fig.~\ref{fig2} together with typical vibrational spectra measured at the ambient pressure. 

The measurement of levitated object is performed using different methods including quadrant or interferometric detection of  reflected laser light, camera video analysis from three directions, or direct optical detection of laser light reflected off the surface below the levitated object casting a shadow of the object on a single-pixel detector (non-interferometric single-pixel detection). The measurement of vibration spectrum of a quartz cube of 0.5~mm side length (see Fig.~\ref{fig2}) reveal the main vibrational frequencies, in agreement with calculated values from the potentials in Fig.~\ref{fig1}. Below we focus on the quartz cube as the levitated object due to its ease of access to various modes, relatively large levitation gaps from the metal surfaces (for reduced eddy current), and our ability to arbitrarily shape and size such particles using a laser cutter.  The spectrum shows two main translational modes along $z$ and $x$ to be respectively at $\sim 10.1$ Hz and $7.3$ Hz. Two distinct lower frequency modes at $0.3$ Hz and $1.4$ Hz are rotation around a horizontal axis and $z$ axis respectively. As mentioned before, the confined rotational mode around $z$ is due to rotational asymmetry in the field profile. 

\subsection*{Dissipation Mechanisms}
In practical settings, the trap is associated with damping that arises from two main mechanisms. At low air pressures \( P \), the residual gas damping rate in the free-molecular regime for a spherical object of cross section $A$ and mass $m$ is approximately given by $\gamma\simeq \alpha P/m$ \cite{Cavalleri_2010}, where $\alpha=A/v_{th}$ is the damping coefficient and $v_{th}$ is the thermal velocity of gas. At $P=10^{-7}$ Torr and including the squeezed film correction, the damping rate of a millimeter-scale sphere only limited by gas damping is expected to be below 0.1~nHz. Achieving such low vacuum levels typically requires annealing of the vacuum chamber. However, caution must be exercised when annealing in the presence of rare-earth magnets, as some of these magnets can lose significant magnetic strength with even modest temperature increases—sometimes as little as 10~°C above room temperature. The is effect is particularly significant in our experiment because magnets at the very center are experiencing very strong magnetic flux not aligned with their initial magnetization direction. 

Eddy currents induced in conductive materials, e.g., magnets' surfaces ( nickel copper nickel coating) or support structures (metallic disk and rod used for field confinement) can also contribute to energy dissipation. For simplicity, consider a spherical dielectric particle of volume \( V \) levitated a distance \( d \) above a magnet, where the magnetic field is confined to a cylindrical region of radius \( a \) directly above the magnet. The vertical damping rate due to eddy currents can be approximated \cite{matsko_mechanical_2020} by $ 2\gamma_z = 2 \frac{V}{\pi a^3} \frac{d}{l_z} \frac{\chi_m g}{l_z},$ where \( l_z \) is the magnetic field penetration depth and \( \chi_m \) is the diamagnetic susceptibility. For typical metallic surfaces (e.g., iron) surrounding the levitated particle to focus the magnetic field, the eddy current damping rate is estimated to be on the order of $5 \times 10^{-4}$ Hz. In our experiment, we use Permendur, which has a resistivity approximately 20 times higher than that of iron, to reduce eddy current losses (i.e., increase \( l_z \)). With this material, we estimate a minimum eddy-current-limited damping rate of about $7 \times 10^{-7} $ Hz for the lowest mode (rotation around $z$ axis) of the cube. The eddy current can further be suppressed by reducing the metal surfaces and their geometries \cite{Zhu:23} (e.g. using laminated steel layers). We note that for the hemisphere, the symmetry of the object suppresses eddy currents, allowing dissipation in the rotational mode about the vertical axis to approach the air-damping limit ($~$nHz at pressures obtain here), an absence of mode coupling.

\begin{figure*}
    \centering
    \includegraphics[width=0.95\linewidth]{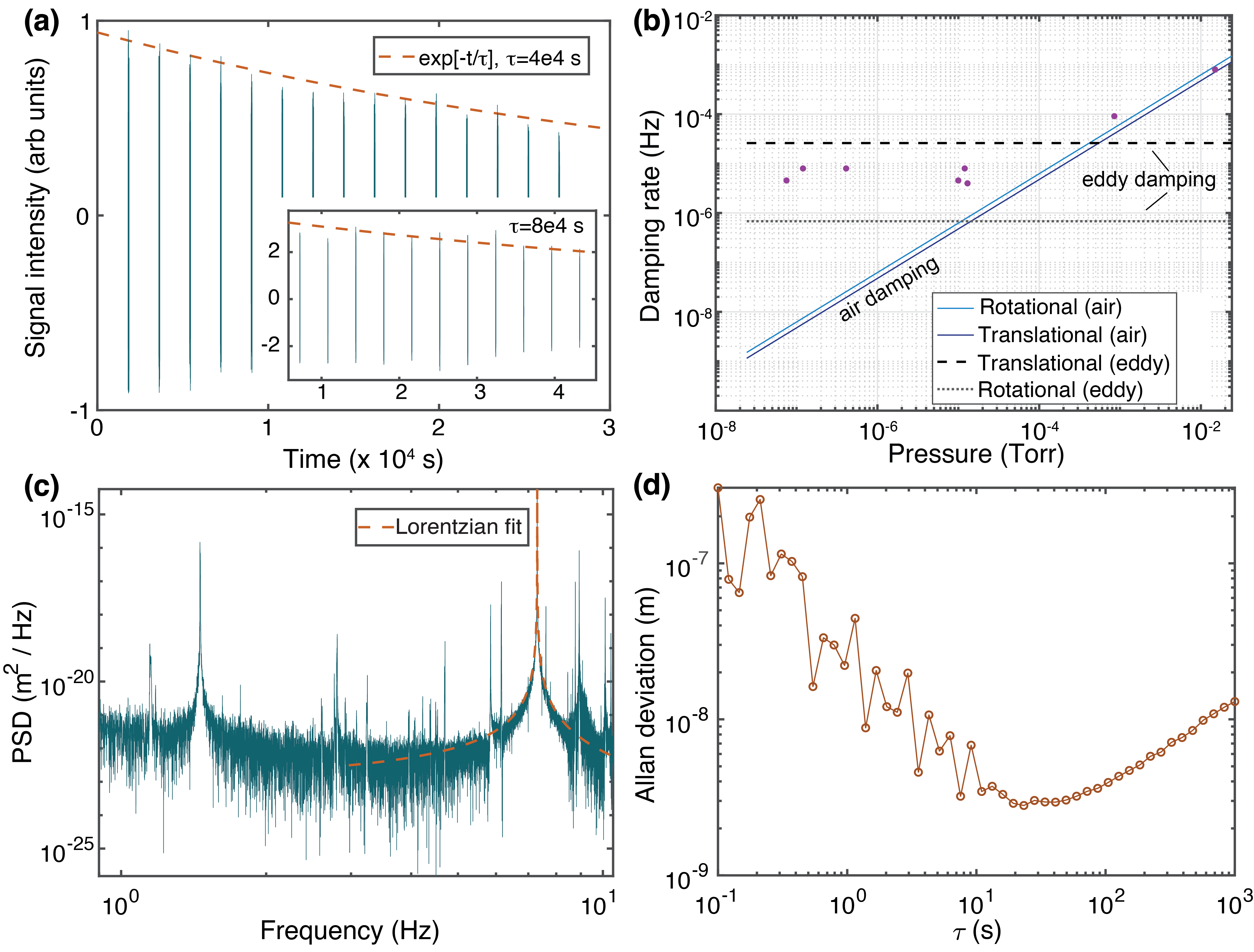}
    \caption{(a) Ring down measurement of vibrational amplitude noise after mechanical excitation of  levitated cube's modes. The data was obtained from a side camera detecting laser scattered light and analyzed similar to a quadrant detector. Inset shows another example of ring down measurement where a long time constant can be inferred. (b) Damping rate measured for several vacuum (circle symbols). The solid lines are the theoretical expectation of air damping in the free-molecular regime for two different modes of vibration (translation and rotation). The effect of squeezed-film damping is negligible. The horizontal lines are the estimated eddy current damping limit for two different modes. (c) Power spectral density (PSD) of levitated quartz cube calibrated using the equipartition theorem. The dashed line is a Lorentzian fit with FWHM of $\sim$0.6~mHz limited by the resolution bandwidth from 2000s-long data. (d) Allan deviation of results in (c) showing a plateau region near 50~s of integration time ($\tau$).   }
    \label{fig3}
\end{figure*}

Fig.~\ref{fig3} (a)-(b) show the measurement result of  damping or dissipation rate at different vacuum pressures. The ring down measurement was obtain after the particle was mechanically or magnetically (using a coil above and at the center of the disk) excited. The ring down  result does not always show a clear decay and in some cases leads to long-term oscillation as shown in the inset of Fig.~\ref{fig3} (a). This is rooted in the nonlinear mode coupling and FPUT recurrence as discussed below. Results indicate that at pressures below $10^{-4}$~Torr, the air damping is insignificant and the primary source of damping is eddy current. The longest decay time measured is $8 \times 10^{4}$~s, limited primarily by the accuracy of amplitude decay measurements in the presence of nonlinear mode coupling and recurrences (see below). We expect the decay time for the lowest vibrational mode to be around $10^{6}$~s, at an elevated (by 50\%) levitation height, which can be achieved by slightly stronger magnets (see Methods).

\subsection*{Displacement and Acceleration Sensitivity}
The acceleration sensitivity is proportional to square root of the ratio between the dissipation rate and mass, $\gamma/m$. The lowest values of this ratio have been reported for suspended mirrors \cite{PhysRevLett.124.221102}, diamagnetic levitation systems \cite{chen_diamagnetic_2022,leng_mechanical_2021}, and nano-particle Paul traps \cite{PhysRevLett.132.133602}, with reported values ranging from $10^{10}$ for nano-particle Paul traps \cite{PhysRevLett.132.133602} to less than $100$ for composite graphite in diamagnetic traps \cite{chen_diamagnetic_2022, tian_feedback_2024}. In our experiment, we achieve $\gamma/m \leq 84$, surpassing cryogenic traps \cite{leng_mechanical_2021,Vinante:2020aa}.

The vibrational spectra can be calibrated using the equipartition theorem to find displacement ($S_{x}$) and force ($S_{F}$) sensitivity, where $S_X=S_F/m\sqrt{(\omega^2-\omega_0^2)^2+\gamma^2\omega^2}$. At vacuum pressures of $8\times10^{-6}$ Torr, the fitted linewidth (limited by integration time and frequency drift) obtained was $6\times10^{-4}$~Hz (see Fig.~\ref{fig3} (c)). We observe a displacement noise floor of $3\times10^{-12}~\text{m}/\sqrt{\text{Hz}}$. From the ring down measurement, we have observed a dissipation rate (full-width half maximum) of $\gamma/2\pi\simeq \SI{4}{\micro\hertz}$ which is insensitive to the frequency drift (see below). We predict a rate of $<\SI{1}{\micro\hertz}$  after re-magnetization of the magnets (see Methods).  At $\SI{1}{\micro\hertz}$, the best gravitational acceleration sensitivity that can be reached (at room temperature) is  $\sqrt{4\gamma k_B T/m}\simeq 62~\text{pg}/\sqrt{\text{Hz}}$.     
Application of the Allan deviation revealed a clear plateau near an averaging time of approximately 50 seconds (see Fig.~\ref{fig3} (d)). This plateau corresponds to the timescale where the drift and correlated noise begin to dominate the stability. 

\subsection*{Nonlinear dynamics}

\begin{figure*}
    \centering
\includegraphics[width=1\linewidth]{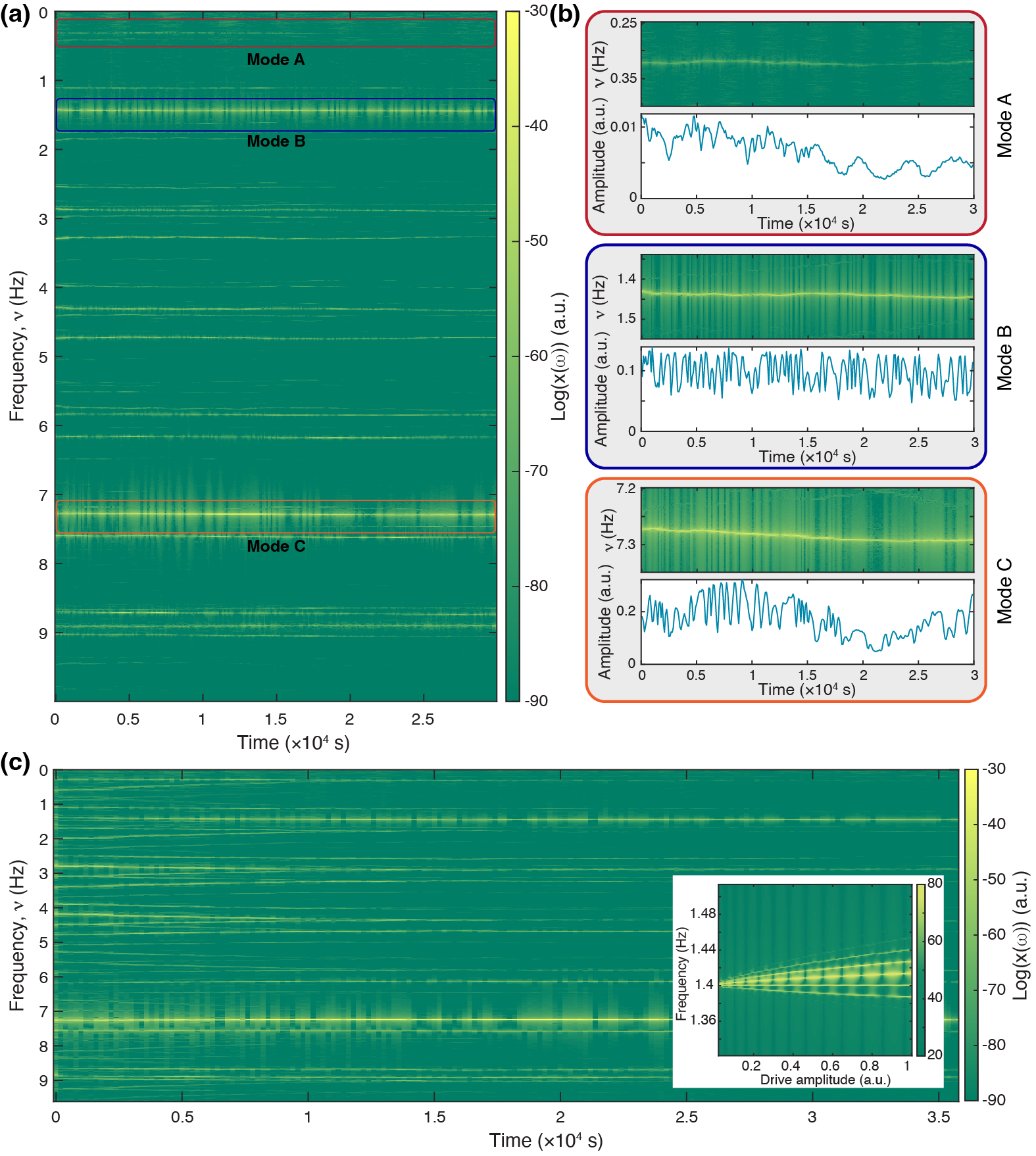}
    \caption{(a) Thermally excited vibrational noise amplitude spectrum ($x(\omega)$) plotted over several hours. (b) The zoomed-in spectra for three main modes (A, B, and C as shown in (a))  and integrated amplitude over each frequency window are plotted as a function of time. (c) Another example of the vibrational spectra recorded after initial excitations to perform ring down measurement. Strong initial drive gives rise to a modulated and shifting spectra. Inset shows nonlinear theory prediction of the fundamental mode behavior under various excitation strength (leading to modulation akin to experimental observations). An arbitrary cubic nonlinearity of $\beta=1$~s$^{-2}$m$^{-2}$ is considered leading to creation of sidebands and frequency shift around the resonance, for a given excitation strength. Vacuum pressure for (a) and (c) was $1.8\times10^{-7}$ and $4.1\times10^{-7}$ Torr, respectively.}
    \label{fig4}
\end{figure*}

The absence of mechanical clamping and the ability to tune trap stiffness and achieve low damping enable clean access to the nonlinear regime. Nonlinear coupling arises naturally from geometric confinement, magnetic field gradients, and particle asymmetry. For a system with generalized coordinates \( x_i \) (corresponding to different modes), the dynamics can be described by coupled differential equations:

\[
\ddot{x}_i + \omega_i^2 x_i + \sum_{j,k} \alpha_{ijk} x_j x_k + \sum_{j,k,l} \beta_{ijkl} x_j x_k x_l + \cdots = 0
\]

where \( \omega_i \) are the linear resonance frequencies, \( \alpha_{ijk} \) and \( \beta_{ijkl} \) represent the quadratic and cubic nonlinear couplings of different vibrational mode, respectively. Quadratic and cubic nonlinearities result in mode coupling, frequency shifts (e.g., Duffing behavior), harmonic generation, and inter-modal energy transfer. Depending on the symmetry and boundary conditions, energy initially localized in one mode can be transferred to others over time, potentially leading to equipartition or recurrence (FPUT physics). FPUT observed that rather than approaching thermal equilibrium, energy initially placed in a single mode undergoes quasiperiodic recurrence, a phenomenon now understood in the context of near-integrability, nonlinear normal modes (NNMs), and Kolmogorov-Arnold-Moser (KAM) theory \cite{Arnold2009}.

To model these systems, a simplified set of coupled Duffing oscillators can be considered:

\[
\ddot{x}_i + \gamma_i \dot{x}_i + \omega_i^2 x_i + \sum_j k_{ij} x_i x_j + \lambda_i x_i^3 = F_i(t)
\]

Here, \( k_{ij} \) models mode-mode coupling, \( \lambda_i \) represents self-nonlinearity, and \( F_i(t) \) is an external drive. Numerical integration can reveal energy redistribution, frequency shifts, and recurrence indicative of FPUT behavior.


Using time-domain displacement measurements, we identify multiple mechanical modes with frequencies below 11 Hz (see Fig.~\ref{fig4} (a)). When investigating the mode more closely, we observe a slow drift in resonant frequency of different modes. Fig.~\ref{fig4} (b) shows mode dynamics for the three main vibrational modes of the system. We have recorded the temperature, background magnetic field, optical power and vacuum pressure over the course of the measurement and observed no correlations between these experimental parameters and the frequency changes observed.  


Spectral analysis reveals evidence of nonlinear mode coupling, manifested partly in the generation of higher harmonics and the appearance of long-lived oscillations in non-fundamental modes. Fig.~\ref{fig4} (c) shows a vibrational spectra under strong initial drive where sideband and frequency shift are evident at early times after the external drive is stopped. 

\begin{figure*}
    \centering
    \includegraphics[width=0.9\linewidth]{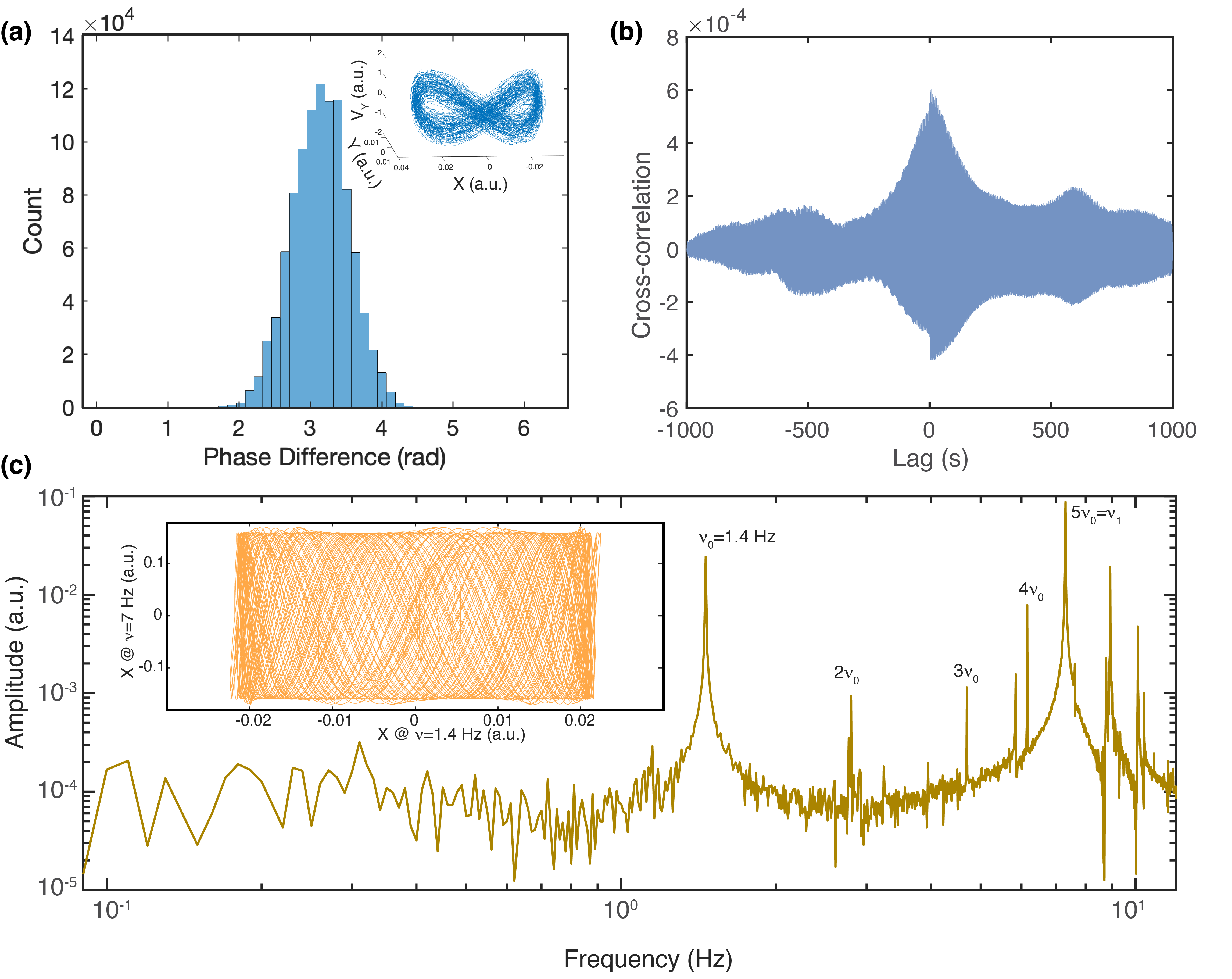}
    \caption{(a) Histogram of phase difference between the fundamental and the 2nd harmonic mode indicates coherence between higher harmonics. Inset shows the 3D phase space plot of position of the two modes, $X$ and $Y$ (fundamental and 2nd harmonic) and velocity of the 2nd harmonic, $V_Y$. (b) Cross correlation of amplitude of two modes shows a peak near zero time lag, another indication of coherence between the two modes. (c) Enhanced parametric mode coupling is observed when a higher harmonic of one mode matches another mode's frequency, in this case the 5th harmonic of 1.4~Hz coincides with $\nu_x$. Inset shows the phase space plot of the two amplitudes (fundamental and higher harmonic modes) where the Lissajous-like figures indicates phase locked motion. Vacuum pressure used was $1.3\times10^{-5}$ Torr.}
    \label{fig5}
\end{figure*}
Phase analysis using the Hilbert transform uncovers partial phase coherence (see Fig.~\ref{fig5} (a)) between the fundamental and its higher harmonic, with the histogram of phase differences peaking near $\pi$, suggesting that energy transfer between modes retains a degree of coordinated timing. Cross-correlation analysis of the time series supports this view (see Fig.~\ref{fig5} (b)), showing oscillatory features in intermodal interactions and delayed thermalization.

As seen in Fig.~\ref{fig5} (c), in certain experimental conditions such as vacuum pressure (leading to drift of particle in the trap), we observe that the fifth harmonic of $\omega_{\theta}$ mode matches the $\omega_x$ mode's frequency. Under this condition we observe enhanced mode coupling where the low-frequency environmental excitations (near $\omega_{\theta}$) gives rise to persistent energy deposited and confined in the system over many hours.   

We examined their trajectories in a two-dimensional phase-space projection. By plotting the displacement of the $x-$ mode ( matching te 5th harmonics) against the fundamental mode ($\nu_{\phi_z}$), we observed evolving Lissajous-like figures (see inset of Fig.~\ref{fig5} (c)). These patterns indicate phase-locked motion at short timescales, but exhibit slow distortions and rotations over longer durations, consistent with weak nonlinear coupling and gradual energy exchange between the modes. The distortion of the classical Lissajous figures is a signature of underlying mode coupling and nonlinearity, leading to a slow drift in relative phase and amplitude. Such evolving trajectories provide visual evidence of the anharmonic and weakly interacting nature of the vibrational modes, which are expected in high-Q macroscopic levitated systems where even subtle nonlinear effects are preserved over long coherence times.

\begin{figure*}
    \centering
    \includegraphics[width=1\linewidth]{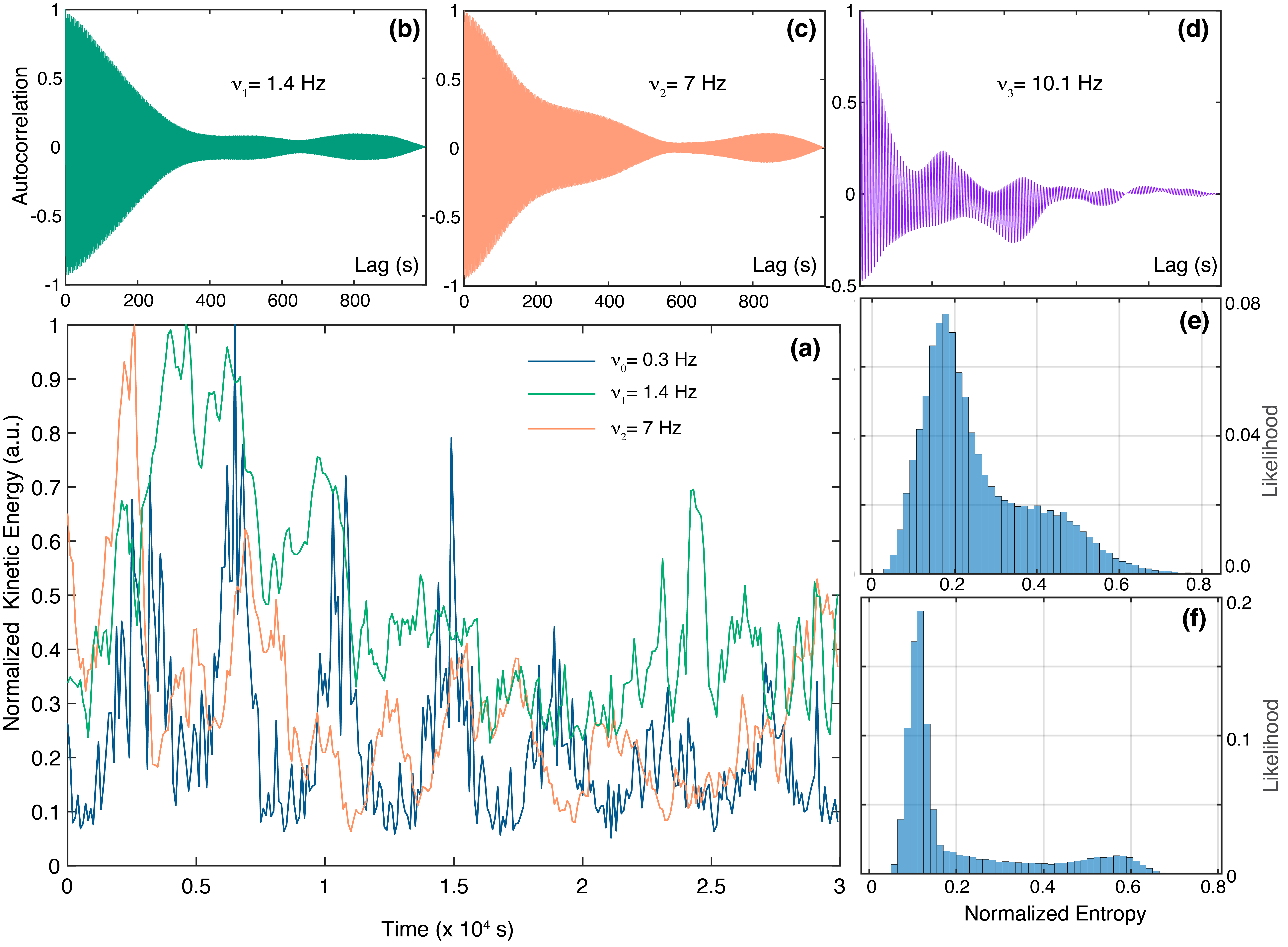}
    \caption{(a) Normalized kinetic energy of three main vibrational modes extracted using digital filtering shows oscillation and exchange of energy between the modes, akin to FPUT recurrence. Vacuum pressure used was $1\times10^{-5}$ Torr. (b)-(d) Autocorrelation of strongly coupled modes at 1.4~Hz and 7~Hz and weakly coupled mode at 10.1~Hz, respectively. (e)-(f) Histogram of spectral entropy of first 10 modes for time windows immediately and 10 hours after the excitation stops, respectively. }
    \label{fig6}
\end{figure*}

FPUT physics has been traditionally studied in optical systems \cite{PhysRevX.8.041017, PhysRevX.4.011054} and cold atoms \cite{kinoshita2006quantum}. The theoretical investigation of FPUT physics in graphene has been suggested \cite{PhysRevLett.112.145503} but experimental studies in mechanical systems is rare. 
 In our system, we are able to observe temporal recurrence in the amplitude evolution of individual modes (see Fig.~\ref{fig6} (a)), indicating non-ergodic, structured energy exchange that deviates from simple thermalization. These features are hallmarks of weakly nonlinear dynamics, consistent with the onset of FPUT-like behavior.

We compute the autocorrelation functions of individual mode amplitudes to probe the temporal coherence and memory effects in the system (see Fig.~\ref{fig6} (b)-(d)) . For the first and second modes, the autocorrelation exhibits a pronounced peak and decays slowly, indicating persistent correlations over long timescales. This is characteristic of low-frequency modes retaining coherence and being less susceptible to rapid energy dispersion. In contrast, the autocorrelation of higher frequency mode $\nu_z$ weakly coupled to others also peaks at zero lag time but shows a narrower profile. This suggests that while this mode is phase-correlated with the base mode, their coherence decays more rapidly, and they exhibit more localized temporal structure. 

In addition, spectral entropy can be used to look for oscillations that can signal periodic energy spreading and refocusing. To quantify this, we compute the instantaneous Shannon entropy of the mode energy distribution at each time step. Let $E_i(t)$ denote the instantaneous energy of the $i$-th mode, and let $P_i(t) = E_i(t) / \sum_j E_j(t)$ be the normalized energy fraction of mode $i$. The entropy at time $t$ is then given by $S(t) = -\sum_i P_i(t) \log P_i(t)$. We normalize $S(t)$ by the maximum possible entropy $S_{\mathrm{max}} = \log(N)$, where $N$ is the total number of modes, so that the normalized entropy $\tilde{S}(t) = S(t)/S_{\mathrm{max}}$ ranges from 0 (perfect energy localization) to 1 (complete equipartition). The histogram of the normalized entropy reveals that after a long evolution time the system spends a significant fraction of its  time in highly ordered states, with entropy values clustered near zero(see Fig.~\ref{fig6} (e)-(f)). As time evolves, nonlinear coupling and dispersion redistribute energy, leading to partial localization in mode space or the formation of coherent structures. The energy spectrum becomes more uneven — a few modes carry most of the energy while others are depleted - reducing the entropy toward smaller values. This indicates that the energy remains predominantly localized in a few modes rather than being evenly distributed, which is characteristic of non-ergodic or weakly thermalizing behavior. 
\begin{figure}
    \centering
    \includegraphics[width=0.7\linewidth]{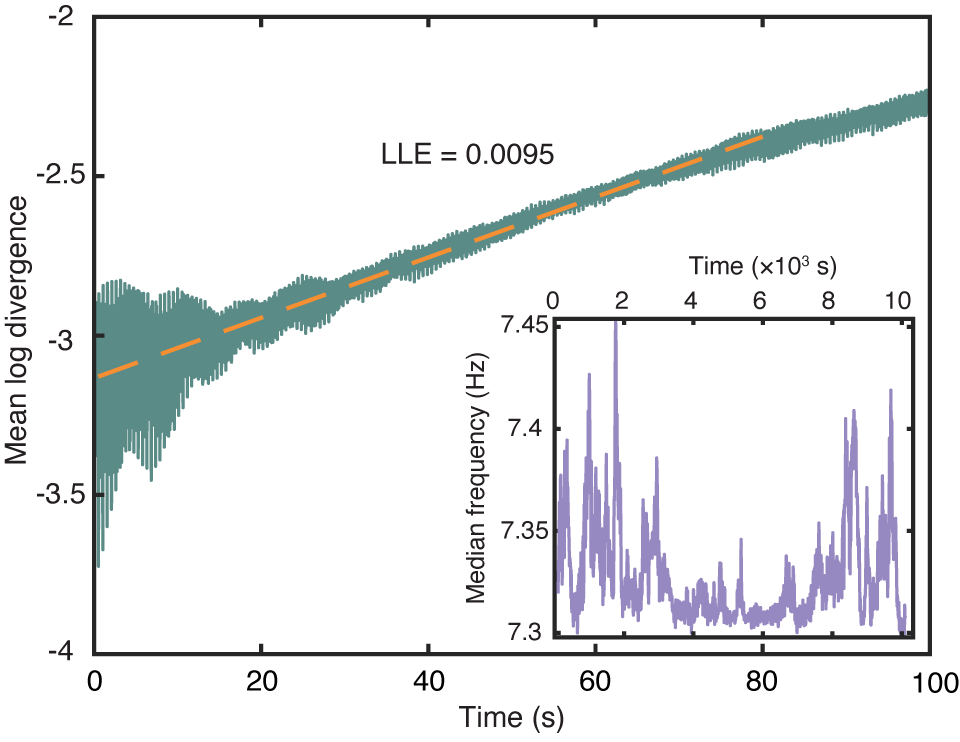}
    \caption{Mean log divergence of nearby phase space trajectories as a function of time, estimated from the reconstructed phase space of the measured time series using embedding dimension 8, time delay 0.5~s, and theiler window of 200~s. The divergence shows a clear linear growth region over 100 seconds, with a fitted slope corresponding to a largest Lyapunov exponent $\lambda_{max}\simeq 0.0095$. Inset shows the median frequency of the 7~Hz mode as a function of time obtained from 30s-long data segments. }
    \label{Fig7}
\end{figure}

  To further study the onset of chatioc behavior, one can estimate the largest Lyapunov exponent (LLE), denoted by \(\lambda_{\max}\), which quantifies the average exponential rate at which two initially close trajectories in a dynamical system diverge in phase space. Given two trajectories starting infinitesimally close with initial separation \(\delta \mathbf{x}_0\), their separation \(\delta \mathbf{x}(t)\) evolves approximately as
\[
\|\delta \mathbf{x}(t)\| \approx \|\delta \mathbf{x}_0\| e^{\lambda_{\max} t}.
\]
A positive value of \(\lambda_{\max} > 0\) indicates sensitive dependence on initial conditions, a hallmark of chaotic dynamics, whereas \(\lambda_{\max} \leq 0\) corresponds to stable or periodic behavior.

  As shown in Fig.~\ref{Fig7}, we estimated the LLE from the measured time-series data using phase-space reconstruction and the method
of nearest-neighbor divergence. The embedding dimension was systematically increased to test the robustness of the result with LLE slope remained essentially constant at $\lambda_{\max}\simeq 0.0095$. The presence of a clear linear divergence region over approximately 100~s, combined with the stability of the slope across embeddings, supports the interpretation of $\lambda_{max}>0$ as an indicator for onset of deterministic chaos.
 
  We also observe discrete jumps in the 7~Hz mode's instantaneous frequency (see inset of Fig.~\ref{Fig7}).  The observed sudden frequency jumps reflects the complex nonlinear dynamics underlying the system signifying transitions towards non-integrable dynamics and the onset of chaotic behavior. The results are consistent with the theoretical framework of KAM and associated nonlinear dynamics.
  
The experimental platform presented here establishes macroscopic diamagnetic levitation as a uniquely low-dissipation, controllable system for exploring nonlinear dynamics, sensing applications, and fundamental physics. Several avenues can be pursued to further enhance performance and broaden scientific relevance.

Although our current system demonstrates exceptionally low mechanical dissipation, further reductions appear within reach. A simple re-magnetization of the magnets can increase the trapping height leading to four fold improvement (see Methods) to reach dissipation rates about \SI{1}{\micro\hertz}. Eddy current damping—arising from field-induced currents in nearby conductive materials—can be mitigated by substituting metallic components with layered or patterned high-resistivity materials and carefully engineering magnetic field confinement. Simulations indicate that for certain modes an optimized geometries may reduce dissipation rates to the nanohertz regime, approaching the fundamental limit set by residual gas damping. This would mark an unprecedented level of isolation in a room-temperature mechanical system.

In the low-dissipation regime and with reduced nonlinearities, the system could approach force sensitivities on the order of few  $pg/\sqrt{\text{Hz}}$. This performance exceeds that of state-of-the-art atomic interferometers and nanomechanical resonators, but in a much simpler, room-temperature setting. Furthermore, by inducing and stabilizing rotational motion of the levitated particle, the system could serve as an ultra-low-power gyroscope \cite{chen2025levitatedmacroscopicrotors10} or accelerometer, potentially useful in portable or space-based inertial navigation systems.

In preliminary experiments, we have shown that the particle can be electrically driven (magnetically using a coil above the disk or electrostatically using voltages applied to the disk and the magnets) with high precision. This same interface could be used to implement feedback cooling \cite{tian_feedback_2024}, dynamically reducing the effective temperature and suppressing unwanted nonlinear effects, or alternatively to enhance specific nonlinearities for controlled studies of energy transfer and chaos.

Moreover, the nonlinear characteristics of the trap—such as the relative strengths of quadratic and cubic terms in the restoring force can be finely tuned by adjusting the magnetic field landscape. This opens the door to engineering Duffing-like or parametrically coupled oscillators, where energy exchange between modes can be selectively enhanced or suppressed. Independent control of these terms would enable detailed studies of bifurcations, synchronization, and dynamical phase transitions in classical systems.

Nonlinear mode coupling, together with high spectral purity, can be a model for frequency comb generation in mechanical resonators \cite{de_Jong_2023,PhysRevX.12.041019}—a phenomenon extensively studied in  \cite{fortier201920} but not yet realized in large-scale mechanical systems. The ability to generate and control mechanical frequency combs could have  important implications for sensing.

On the other hand, because the levitated object is dielectric, the system can be adapted to incorporate optical trapping forces. As previously proposed \cite{PhysRevLett.111.183001,jiang2020superconducting}, it is possible to levitate a dielectric mirror and use it as the end-mirror of an optical cavity. Combining diamagnetic and ``coherent" optical trapping could yield hybrid traps with greatly enhanced stiffness and mechanical frequencies, allowing access to the ultra-high-$Q$ regime required for many sensing applications as well as tests of gravitational decoherence \cite{PhysRevLett.119.240401}, the Schrödinger-Newton equation \cite{PhysRevD.96.044008, PhysRevD.93.124049, PhysRevD.93.096003}, and other collapse models in quantum mechanics. Such a system would enable sensitive searches for deviations from quantum linearity at macroscopic scales.


 The platform studied here offers high tunability, e.g., via particle shape, magnetic trap geometry, or external perturbation, enabling accurate sensing and controlled studies of: KAM transitions, resonance overlap and chaos, prethermal plateaus, weak turbulence and thermalization scaling laws. 
This work lays the foundation for a new experimental paradigm in classical nonlinear physics—one where macroscopic coherence, strong nonlinearity, and ultra-low damping coexist, opening a path to development of novel multimodal sensors and study of nonlinear phenomena long considered purely theoretical and numerical.

\section*{Methods}
Quartz cubes were cut from a 2-inch quartz wafer (0.5~mm thick) using a CNC laser cutter (LPKF ProtoLaser R). To stabilize the wafer during the cutting process and prevent suction from the laser cutter, the quartz wafer was mounted on a 4-inch silicon wafer using a PMMA adhesive layer. The cutting was performed at a laser frequency of 300 kHz with a power of 7.3 W. To maintain vertical sidewalls and ensure precise beam focusing, the laser focus offset was adjusted every \SI{250}{\micro\meter} of cube height. At each focus offset, 200 repetitions were applied using the specified parameters. Following the cutting process, PMMA was removed with acetone, and the quartz cubes were separated from the wafer.

Illumination of the particle for measurement was provided by an 800 nm low-noise Ti:Sapphire laser. A thin reflective glass surface placed beneath the levitated cube projected its shadow onto a single-pixel detector. Most data were collected overnight during weekends, typically over 10-hour intervals. Two cameras, positioned above and to the side of the levitated particle, recorded the scattered light. The resulting images were analyzed to extract various vibrational modes and compared with theoretical predictions. Because this method is largely insensitive to alignment and power drift, it was used for ring-down measurements. Additional detection techniques, including interferometric and quadrant photodiode methods, were employed to validate the sensitivity and signal-to-noise ratio of the single-pixel and camera-based approaches.

To verify that particle frequency and amplitude fluctuations were not influenced by environmental factors, we monitored variation of the magnetic field ($\Delta B$) a few centimeters above the magnet, outside the chamber, temperature ($\Delta T$) at three locations adjacent to the chamber, air flow near the optical table, laser power, and typical vacuum pressure ($P$). Although variations were observed, e.g. $\Delta B\simeq0.7$~G, $\Delta T\simeq1.1^{\circ}$~C, $ P= 1.27(\pm0.01)\times 10^{-5}$~Torr, over a period of 10~h in these environmental parameters, no correlation with fast particle dynamics ($<1$~h time scale) was found. The slow frequency drift over long time scales (several hours), however, may be attributed to the slow drift of the background magnetic field.  

For particle loading, a vacuum tweezer was used to release the particle into the central hole of the top disk. A small mechanical perturbation caused the particle to “jump” into the trapping location. Its height and position relative to the magnet center were fine-tuned using three screws to adjust the tilt and height of the top disk. The entire setup was mounted on adjustable legs to allow coarse control of the overall alignment.  

To reach low vacuum pressures, we slightly annealed the chamber at about 20 degrees above the room temperature. This led to slight demagnetization of the rare-earth magnets lowering the trapping height of the particle. Simple re-magnetization of the magnets will regain the higher trapping height leading to reduction of the eddy current damping by a factor of three to four for all vibrational modes, with $\nu_z$ having the lowest dissipation.

\section*{Data availability}
The data supporting the findings of this study are available upon reasonable request.

\section*{Acknowledgments}
 We thank Andrew Geraci and Selim Shahriar for their valuable discussions and comments. This work was supported by DARPA Young Investigator Award, Grant No. D24AP00326-00.
 
\bibliography{sn-bibliography/sn_bibliography}

\end{document}